\documentclass[conference]{IEEEtran}
\ifCLASSINFOpdf
  \usepackage[pdftex]{graphicx}
  \DeclareGraphicsExtensions{.pdf,.jpeg,.png}
\else
\fi
\usepackage[cmex10]{amsmath}
\usepackage{amsfonts}  

\begin{document}
%
\title{Scalar-linear Solvability of Matroidal Networks Associated with Representable Matroids}

\author{\IEEEauthorblockN{Anthony Kim and Muriel M\'{e}dard}
\IEEEauthorblockA{Research Laboratory of Electronics\\
Massachusetts Institute of Technology\\
Cambridge, MA 02139, USA\\
Email: \{tonyekim, medard\}@mit.edu}}
\newtheorem{theorem}{Theorem}
\newtheorem{corollary}[theorem]{Corollary}
\newtheorem{lemma}[theorem]{Lemma}
\newtheorem{observation}[theorem]{Observation}
\newtheorem{proposition}[theorem]{Proposition}
\newtheorem{definition}[theorem]{Definition}
\newtheorem{claim}[theorem]{Claim}
\newtheorem{fact}[theorem]{Fact}
\newtheorem{assumption}[theorem]{Assumption}
\newtheorem{remark}[theorem]{Remark}
\newtheorem{example}[theorem]{Example}
\newtheorem{conjecture}[theorem]{Conjecture}
\numberwithin{equation}{section}   

\newcommand{\CC}{\mathbb C} 
\newcommand{\RR}{\mathbb R}
\newcommand{\ZZ}{\mathbb Z}
\newcommand{\QQ}{\mathbb Q}
\newcommand{\NN}{\mathbb N}
\newcommand{\FF}{\mathbb F}
\newcommand{\GGG}{\mathcal G}
\newcommand{\XXX}{\mathcal X}
\newcommand{\BBB}{\mathcal B}
\newcommand{\CCC}{\mathcal C}
\newcommand{\EEE}{\mathcal E}
\newcommand{\FFF}{\mathcal F}
\newcommand{\MMM}{\mathcal M}
\newcommand{\NNN}{\mathcal N}
\newcommand{\III}{\mathcal I}
\newcommand{\AAA}{\mathcal A}
\newcommand{\SSS}{\mathcal S}
\newcommand{\sign}{\operatorname{sign}}
\newcommand{\ideal}{\operatorname{Ideal}}
\newcommand{\In}{\operatorname{In}}
\newcommand{\Out}{\operatorname{Out}}
\newcommand{\Span}{\operatorname{span}}
\newcommand{\vect}[1]{\vec{\boldsymbol{#1}}}

\maketitle

\begin{abstract}
We study matroidal networks introduced by Dougherty et al. We prove the converse of the following theorem: If a network is scalar-linearly solvable over some finite field, then the network is a matroidal network associated with a representable matroid over a finite field. It follows that a network is scalar-linearly solvable if and only if the network is a matroidal network associated with a representable matroid over a finite field. We note that this result combined with the construction method due to Dougherty et al. gives a method for generating scalar-linearly solvable networks. Using the converse implicitly, we demonstrate scalar-linear solvability of two classes of matroidal networks: networks constructed from uniform matroids and those constructed from graphic matroids. 
\end{abstract}


%
\IEEEpeerreviewmaketitle

\section{Introduction}
In 2000, Ahlswede et al.~\cite{ahlswede:network} introduced the network coding scheme to the problem of communicating information in networks by allowing intermediate nodes to code on the incoming packets. They showed that the extended capability of intermediate nodes gives greater information throughput than in the traditional routing scheme. They also showed that the capacity of a multicast network is equal to the minimum of min-cuts between source and receiver nodes. 

Li et al.~\cite{li:linear} showed that linear network coding is sufficient for multicast networks. Koetter and M\'{e}dard~\cite{koetter:algebra} reduced the problem of determining scalar-linear solvability to solving a set of polynomial equations over some finite field and suggested connections between scalar-linearly solvable networks and nonempty varieties in algebraic geometry. They showed that many special case networks, such as two-level multicast and disjoint multicast, can be shown to be scalar-linearly solvable by their method. Dougherty et al.~\cite{dougherty:poly} strengthened the connection by demonstrating solvably equivalent pairs of networks and polynomial collections; for any polynomial collection, there exists a network that is scalar-linearly solvable over field $F$ if and only if the polynomial collection is solvable over $F$. It is known that scalar-linear network codes are not sufficient in general. The M-network due to Koetter in \cite{medard:mnetwork} is a network with no scalar-linear solution but has a vector-linear solution. Lehman and Lehman~\cite{lehman} using 3-CNF formulas also provided an example where a vector solution is necessary.

Dougherty et al.~\cite{dougherty:matroid} introduced matroidal networks and suggested connections between networks and matroids. They used matroidal networks constructed from well-known matroids to show  in \cite{dougherty:insuff} that not all solvable networks have a linear solution over some finite-field alphabet and vector dimension. They also constructed a matroidal network to show that Shannon-type information inequalities are not sufficient for computing network coding capacities in general. Recently, El Rouayheb et al.~\cite{rouayheb:matroid} strengthened the connection between networks and matroids by constructing ``solvably equivalent'' pairs of networks and matroids; the network has a vector-linear solution over a field if and only if the matroid has a multilinear representation over the same field.

In this paper, we further study the matroidal networks. In particular, we prove the converse of a theorem by Dougherty et al.~\cite{dougherty:matroid} which states that if a network is scalar-linearly solvable then it is a matroidal network associated with a representable matroid over a finite field. It follows that a network is scalar-linearly solvable if and only if it is a matroidal network associated with a representable matroid over a finite field. We use the result implicitly to show that matroidal networks constructed from uniform and graphic matroids are scalar-linearly solvable. The main idea is to construct a scalar-linear network code as we construct such a matroidal network so as to satisfy all the connections in the resulting network. Our results give a set of scalar-linearly solvable networks that might be different from those we already know. 

The paper is organized as follows. In Section II, we give a network coding model. In Section III, we define matroids and three classes of matroids: uniform, graphic, and representable matroids. In Section IV, we define matroidal networks and provide the construction method due to Dougherty et al.~\cite{dougherty:matroid}. In Section V, we prove the converse of the theorem by Dougherty et al. and demonstrate scalar-linear solvability of matroidal networks constructed from uniform and graphic matroids. Finally, we conclude in Section VI. 

\section{Network Coding}
We give a formulation of scalar-linear network codes; most of it is adapted from \cite{dougherty:matroid}.

A network is a finite, directed, acyclic multigraph with node set $\nu$, edge set $\epsilon$, message set $\mu$, a source mapping $S: \nu \rightarrow 2^\mu$, a receiver mapping $R:\nu\rightarrow 2^\mu$ and an alphabet $\AAA$. For each node $x$, if $S(x)$ is nonempty, then $x$ is called a {\em source} and if $R(x)$ is nonempty then $x$ is called a {\em receiver}. The elements of $S(x)$ are called the {\em messages generated by $x$} and the elements of $R(x)$ are called the {\em messages demanded by $x$}. An alphabet $\AAA$ is a finite set with at least two elements. For each node $x$, let $\In(x)$ denote the set of messages generated by $x$ and in-edges of $x$. Let $\Out(x)$ denote the set of messages demanded by $x$ and out-edges of $x$. 

For each node $x$, we fix an ordering of $\In(x)$ such that all messages in the resulting list occur before the edges. Let $F$ be a finite field. For each edge $e=(x,y)$, an {\em edge function} is a map $f_e : F^\alpha \times F^\beta \rightarrow F$ where $\alpha$ and $\beta$ are the number of messages and in-edges of $x$ respectively. For every $x\in \nu$ and $m\in R(x)$, a {\em decoding function} is a map $f_{x,m}: F^\alpha \times F^\beta \rightarrow F$ where $\alpha$ and $\beta$ are the number of messages and in-edges of $x$. A {\em scalar-linear code} for a network is an assignment of such edge functions and decoding functions that are linear over a finite field $F$ that is sufficiently large enough to represent all elements in $\AAA$, i.e. $\lvert F \rvert \geq \lvert \AAA \rvert$. Each instance of a message is an element from $\AAA$ which is uniquely represented by an element from $F$. A {\em message assignment} is a map $a:\mu \rightarrow F$. We recursively define the function $c: \epsilon \rightarrow F$ as $c(e) = f_e(a(x_1), \ldots, a(x_\alpha), c(x_{\alpha+1}), \ldots, c(x_{\alpha+\beta}))$ for each edge $e=(x,y)$, where $x_1, \ldots, x_\alpha$ are the messages generated by $x$ and $x_{\alpha+1}, \ldots, x_{\alpha+\beta}$ are the in-edges of $x$. 

For a given scalar-linear network code, if for every message assignment $a:\mu \rightarrow F$ we have $f_{x,m}(a(x_1), \ldots, a(x_\alpha), c(x_{\alpha+1}), \ldots, c(x_{\alpha+\beta})) = a(m)$ for receiver $x$ and message $m$ demanded by $x$, then node $x$'s {\em demand $m$ is satisfied}. A scalar-linear network code is a scalar-linear solution if every demand in the network is satisfied.

Given a scalar-linear network code over a finite field $F$, we define the corresponding global network code on edges and node-message pairs to be a mapping to column vectors as $\phi:\epsilon \cup (\nu \times \mu) \rightarrow F^{\lvert \mu \rvert}$ such that
\begin{enumerate}
\item for any edge $e$, $c(e) = (a(m_1), \ldots, a(m_{\lvert \mu \rvert})) \cdot \phi_e$, and
\item for any node $x$ and message $m$, $\phi_{x,m}$ has a 1 in the coordinate corresponding to $m$ if and only if $m \in S(x)$. 
\end{enumerate}
We refer to $\phi_e$ and $\phi_{x,m}$ as {\em global coding vectors}. A global network code is {\em valid} if for every node $x\in \nu$ and out-edge $e$ of $x$, $\phi_e \in \Span(\{\phi_y \: : \: y\in \In(x)\})$, where if $y$ is a message $m$ generated by $x$ then $\phi_y$ is understood to be $\phi_{x,m}$. For each receiver $x$ and message $m$ demanded by $x$, the $x$'s demand is satisfied if and only if the standard basis vector corresponding to $m$ is in the span of $\{\phi_y \: : \: y\in \In(x)\}$. We say that a global network code is a scalar-linear solution if all the demands are satisfied. Given a valid global network code, we can derive an equivalent scalar-linear network code. Therefore the notion of scalar-linear solution with edge functions and decoding functions is equivalent to the notion of scalar-linear global network code solution. In this paper, we shall mostly use the global network codes.

\section{Matroids}
We define matroids and three classes of matroids. See \cite{oxley:matroid} for more background on matroids.

\begin{definition}
A matroid $\MMM$ is an ordered pair $(\SSS, \III)$ consisting of a set $\SSS$ and a collection $\III$ of subsets of $\SSS$  satisfying the following conditions:
	\begin{enumerate}
	\item $\emptyset \in \III$;
	\item If $I \in \III$ and $I'\subseteq I$, then $I'\in \III$;
	\item If $I_1$ and $I_2$ are in $\III$ and $|I_1|<|I_2|$, then there is an element $e$ of $I_2 \setminus I_1$ such that $I_1 \cup \{e\} \in \III$. 
	\end{enumerate}
\end{definition}
The set $\SSS$ is called the {\em ground set} of the matroid $\MMM$. A subset $X$ of $\SSS$ is an {\em independent set} if it is in $\III$; $X$ is a {\em dependent set} if not. A {\em base} $B$ of $\MMM$ is a maximal independent set; for all element $e\in \SSS \setminus B$, $B \cup \{e\} \notin \III$. It can be shown that all bases have the same cardinality. A {\em circuit} of $\MMM$ is a minimal dependent set; for all element $e$ in $C$, $C \setminus \{e\} \in \III$. For each matroid, there is an associated function $r$ called {\em rank} that maps the power set $2^\SSS$ into the set of nonnegative integers. The rank of a set $X \subseteq \SSS$ is the maximum cardinality of independent set contained in $X$. 

\begin{definition}
Two matroids $\MMM_1=(\SSS_1, \III_1)$ and $\MMM_2=(\SSS_2, \III_2)$ are isomorphic if there is a bijection $\psi$ from $\SSS_1$ to $\SSS_2$ such that for all $X \subseteq \SSS_1$, $X$ is independent in $\MMM_1$ if and only if $\psi(X)$ is independent in $\MMM_2$. 
\end{definition}

\begin{definition}[Uniform Matroids]
Let $c,d$ be nonnegative integers such that $c\leq d$. Let $\SSS$ be a $d$-element set and $\III$ be the collection $\{X \subseteq \SSS \: : \: \lvert X \rvert \leq c\}$. We define the uniform matroid of rank $c$ on the $d$-element set to be $U_{c,d} = (\SSS, \III)$.
\end{definition}

\begin{definition}[Graphic Matroids]
Let $G$ be an undirected graph with $\SSS$ the set of edges. Let $\III=\{X\subseteq \SSS \medspace : \medspace X \text{ does not contain a cycle}\}$. We define the graphic matroid associated with $G$ as $\MMM(G) = (\SSS,\III)$.
\end{definition}

\begin{definition}[Representable/Vector Matroid]
Let $A$ be a $d_1 \times d_2$ matrix over some field $F$. Let $\SSS = \{1,\ldots, d_2\}$ where each $i$ corresponds to the $i$th column vector of $A$ and $\III = \{X\subseteq \SSS \medspace : \medspace \text{corresponding column vectors form an independent set} \}$. We define the vector matroid associated with $A$ as $\MMM(A) = (\SSS, \III)$. A matroid $\MMM$ is $F$-representable if it is isomorphic to a vector matroid of some matrix of $D$ over field $F$. A matroid is representable if it is representable over some field. Note that $F$ is not necessarily finite.
\end{definition}

The bases of $U_{c,d}=(\SSS,\III)$ are exactly subsets of $\SSS$ of cardinality $c$ and the circuits are subsets of $\SSS$ of cardinality $c+1$. Each base of $\MMM(G)$ is a spanning forest of $G$, hence an union of spanning trees in connected components of $G$, and each circuit is a single cycle within a connected component. It is known that the graphic matroids are representable over any field $F$. On the other hand, the uniform matroid $U_{2,4}$ is not representable over $GF(2)$.

\section{Matroidal Networks} 
We define matroidal networks and present a method for constructing matroidal networks from matroids; for more details and relevant results, we refer to \cite{dougherty:matroid}. 
\begin{definition}\label{def:matroidal} 
Let $\NNN$ be a network with message set $\mu$, node set $\nu$, and edge set $\epsilon$. Let $\MMM=(\SSS,\III)$ be a matroid with rank function $r$. The network $\NNN$ is a {\em matroidal network} associated with $\MMM$ if there exists a function $f:\mu \cup \epsilon\rightarrow \SSS$, called the {\em network-matroid mapping}, such that the following conditions are satisfied:
\begin{enumerate}
\item $f$ is one-to-one on $\mu$;
\item $f(\mu)\in \III$;
\item $r(f(\In(x))) = r(f(\In(x)\cup \Out(x)))$, for every $x\in \nu$.
\end{enumerate}
\end{definition}

\begin{theorem}[Construction Method]\label{thm:construction}
Let $\MMM=(\SSS, \III)$ be a matroid with rank function $r$. Let $\NNN$ denote the network to be constructed, $\mu$ its message set, $\nu$ its node set, and $\epsilon$ its edge set. Then the following construction method will construct a matroidal network $\NNN$ associated with $\MMM$. 

The construction will simultaneously construct the network $\NNN$, the function $f:\mu\cup\epsilon \rightarrow \SSS$, and an auxiliary function $g:\SSS \rightarrow \nu$, where for each $x\in\SSS$, $g(x)$ is either
\begin{enumerate}
\item a source node with message $m$ and $f(m)=x$; or
\item a node with in-degree 1 and whose in-edge $e$ satisfies $f(e)=x$.
\end{enumerate}
The construction is completed in 4 steps and each step can be completed in potentially many different ways:

{\noindent}\textit{\underline{Step 1}}: Choose any base $B=\{b_1, \ldots, b_{r(\SSS)}\}$ of $\MMM$. Create network source nodes $n_1, \ldots, n_{r(\SSS)}$ and corresponding messages $m_1, \ldots, m_{r(\SSS)}$. Let $f(m_i)=b_i$ and $g(b_i)=n_i$. 

{\noindent}\textit{\underline{Step 2}}: (to be repeated until no longer possible).\\
Find a circuit $\{x_0, \ldots, x_j\}$ in $\MMM$ such that $g(x_1),\ldots, g(x_j)$ have been already defined but not $g(x_0)$. Then we add:
\begin{enumerate}
\item a new node $y$ and edges $e_1, \ldots, e_j$ such that $e_i$ connects $g(x_i)$ to $y$. Let $f(e_i)=x_i$.
\item a new node $n_0$ with a single in-edge $e_0$ that connects $y$ to $n_0$. Let $f(e_0) =x_0$ and $g(x_0) =n_0$. 
\end{enumerate}

{\noindent}\textit{\underline{Step 3}}:(can be repeated arbitrarily many times).\\
If $\{x_0, \ldots, x_j\}$ is a circuit of $\MMM$ and $g(x_0)$ is a source node with message $m_0$, then add to the network a new receiver node $y$ which demands the message $m_0$ and has in-edges $e_1, \ldots, e_j$ where $e_i$ connects $g(x_i)$ to $y$. Let $f(e_i) =x_i$. 

{\noindent}\textit{\underline{Step 4}}:(can be repeated arbitrarily many times).\\
Choose a base $B=\{x_1, \ldots, x_{r(\SSS)}\}$ of $\MMM$ and create a receiver node $y$ that demands all of the network messages and has in-edges $e_1, \ldots, e_{r(\SSS)}$ where $e_i$ connects $g(x_i)$ to $y$. Let $f(e_i)=x_i$. 
\end{theorem}$\phantom{XXX}$

The following theorem is from \cite{dougherty:matroid}. We give the proof for completeness using the network coding model of this paper.
\begin{theorem}\label{thm:scalarmatroidal}
If a network is scalar-linearly solvable over some finite field, then the network is matroidal. Furthermore, the network is associated with a representable matroid over a finite field.
\end{theorem}
\begin{IEEEproof}
Let $\NNN$ denote the network. Fix a scalar-linear solution to the network over a finite field $F$, and let $m_1,\ldots, m_{\lvert \mu \rvert}$ be the network messages. Let $\phi$ be the corresponding global network code. If $m$ is a message, then $\phi_m$ is the standard basis vector corresponding to $m$. If $e$ is an edge, then the symbol carried over edge $e$ is the linear combination $(a(m_1), \ldots, a(m_{\lvert \mu \rvert})) \cdot \phi_e$. Form a matrix $A$ with $\phi_m$ for all message $m$ and $\phi_e$ for all edge $e$ as columns and let $\MMM=(\SSS,\III)$ be the corresponding representable matroid over $F$. We define the network-matroid mapping $f$ from $\mu \cup \epsilon$ to $\SSS$ as the correspondence map. We show that all the three properties of matroidal networks hold. Clearly, $f$ is one-to-one on $\mu$ by construction. For message $m$, the corresponding column of $A$ has 1 in the coordinate corresponding to $m$ and zeros elsewhere. Then $\lvert \mu \rvert$ columns associated with messages are independent, hence $f(\mu) \in \III$. To see the third property, note that for each $e \in \Out(x)$ and $x\in \nu$, $\phi_e$ is a linear combination of coding vectors in $\{\phi_y \: : \: y \in \In(x)\}$. Therefore
\begin{equation*}
r(f(\In(x) \cup \Out(x))) = r(f(\In(x))), \text{ for every $x\in \nu$}.
\end{equation*}
Hence the theorem follows.
\end{IEEEproof}

\section{Scalar-linear Solvability}
We prove the converse of Theorem~\ref{thm:scalarmatroidal} and demonstrate scalar-linear solvability of matroidal networks constructed from uniform and graphic matroids. We note a general method for constructing scalar-linearly solvable networks. We need a technical lemma which we state below without a proof because of the paper length:
\begin{lemma}\label{lem:rep}
Let $A$ be a $d_1 \times d_2$ matrix over a finite field $F$ and $\MMM(A)$ be the corresponding representable matroid. Then there exists an arbitrarily large finite field $F'$ and a $d_1 \times d_2$ matrix $A'$ over $F'$ such that the corresponding matroid $\MMM(A')$ is isomorphic to $\MMM(A)$. 
\end{lemma}

\begin{theorem}\label{thm:converse}
If a network $\NNN$ is matroidal and is associated with a representable matroid over a finite field $F$, then $\NNN$ is scalar-linearly solvable.
\end{theorem}
\begin{IEEEproof}
Let $\NNN$ be the matroidal network with message set $\mu$, node set $\nu$, edge set $\epsilon$, and alphabet $\AAA$. Let $A$ be the $d_1 \times d_2$ matrix over the finite field $F$ such that $\NNN$ is a matroidal network associated with the corresponding matroid $\MMM(A)=(\SSS,\III)$. By Lemma~\ref{lem:rep}, we assume that the finite field is large enough to represent all elements in $\AAA$, i.e. $\lvert F \rvert \geq \lvert \AAA \rvert$. By Definition~\ref{def:matroidal}, there exists a network-matroid mapping $f:\mu \cup \epsilon \rightarrow \SSS$. Assume $r(\SSS) = d_1$; otherwise we remove redundant rows without changing the structure of the matroid. Let $f(\mu)=\{i_1, \ldots, i_{\lvert \mu \rvert}\}$. As $f(\mu)\in \III$, the columns indexed by $f(\mu)$ form an independent set.  We extend $f(\mu)$ to a basis $B$ of $F^{d_1}$ by adding columns of $A$. Wlog, assume the first $d_1$ columns of $A$ form the basis $B$ after reordering. By performing elementary row operations, we uniquely express $A$ in the form
\begin{equation*}
A = [ I_{d_1}\: | \: A']
\end{equation*}
where $A'$ is a $d_1 \times (d_2-d_1)$ matrix and such that $\{i_1, \ldots, i_{\lvert \mu \rvert}\}$ now corresponds to the first $\lvert \mu \rvert$ columns of $A$. Note that the structure of the corresponding matroid stays the same. Let $A_i$ denote the $i$th column of $A$. We introduce dummy messages $m_{\lvert \mu \rvert +1}, \ldots, m_{d_1}$ in $\NNN$ by adding a disconnected node that generates these messages. We assign global coding vectors on the resulting $\NNN$ as follows:
	\begin{enumerate}
	\item for each edge $e$, let $\phi_e = A_{f(e)}$; and 
	\item for node $x$ and message $m$ generated or demanded by $x$, $\phi_{x,m} = A_{f(m)}$.
	\end{enumerate}
We show that the global network code defined above is valid and satisfies all the demands. For each node $x\in \nu$, we have $r(f(\In(x))) = r(f(\In(x) \cup \Out(x)))$. It follows that for every edge $e \in \Out(x)$, $A_{f(e)}$ is a linear combination of $\{ A_{f(e')} \: : \: e'\in \In(x)\}$. Equivalently, $\phi_e$ is a linear combination of $\{\phi_y \: : \: y\in \In(x)\}$, and hence the global network code is valid. For receiver node $x$ and message $m$ demanded by $x$, $m \in \Out(x)$ and $\phi_{x,m}$ is the standard basis vector corresponding to $m$. Since the global network code is valid, it follows that message $m$ can be recovered. Removing the dummy messages, it follows that $\NNN$ is scalar-linearly solvable.
\end{IEEEproof}

Putting Theorems~\ref{thm:scalarmatroidal} and \ref{thm:converse} together, it follows that a network is scalar-linearly solvable if and only if the network is a matroidal network associated with a representable matroid over a finite field. As an illustration of Theorem~\ref{thm:converse}, we demonstrate scalar-linear solvability of matroidal networks constructed from uniform and graphic matroids. The use of the theorem is implicit in the construction of global network codes in the proofs. 

\begin{theorem}\label{thm:uniform}
Any matroidal network constructed from an uniform matroid is scalar-linearly solvable over a sufficiently large finite field of any characteristic.
\end{theorem}
\begin{IEEEproof}
The main idea of the proof is to construct inductively a global network code that satisfies all the demands as we construct the matroidal network $\NNN$. We observe that the natural set of objects to induct on is the set of in-edges of nodes in the range of partially constructed $g$, since all the intermediate and receiver nodes to be created in Steps 2-4 are based on these nodes.

Let $U_{c,d}$ be an uniform matroid. If $c=0$, the network $\NNN$ will not have any messages and hence is trivially scalar-linearly solvable. If $c=1$, there would be exactly one source node generating the only message in $\NNN$, and any receiver node will have a path from the source node. Hence we can simply route the only message to satisfy all the demands and $\NNN$ is scalar-linearly solvable. If $c=d$, then there are no circuits in the matroid and no intermediate nodes in the network $\NNN$. All receiver nodes are created in Step 4 and have an edge from each source node. Clearly, the network $\NNN$ has a routing solution and is scalar-linearly solvable.

We consider the more general case where $c\geq 2$ and $d>c$. Let $p$ be a prime. We choose a finite field $F$ of characteristic $p$ that is sufficiently large enough so that each symbol in the alphabet $\AAA$ can be represented and $\lvert F\rvert \geq \binom{d-1}{c-1}$. Note that $F= \FF_{p^l}$ for some sufficiently large $l$. Since each base has size of $c$, there will be exactly $c$ messages in network $\NNN$. As we construct, we maintain that if there are $k$ nodes in the range of partially constructed $g$ and $\phi_1, \ldots, \phi_k$ are global coding vectors on in-edges of the nodes then 
\begin{equation}\label{prop:u1}
\text{any $c$ vectors of $\phi_{1}, \ldots, \phi_{k}$ forms a basis in $F^c$.}
\end{equation}

In Step 1, we choose a base $B = \{b_1, \ldots, b_c\}$ and create network source nodes $n_1, \ldots, n_c$ and corresponding messages $m_1, \ldots, m_c$. We define $f(m_i)=b_i$ and $g(b_i)=n_i$. We add an auxiliary in-edge $e_i$ to each source $n_i$; this makes the analysis easier without changing the problem. We assign global coding vector $\phi_{n_i}$ to be the $i$th standard basis vector $\tilde{e}_i$. Clearly, $\phi_{n_1}, \phi_{n_2}, \ldots, \phi_{n_c}$ satisfies (\ref{prop:u1}). In Step 2, we repeatedly find a circuit $\{x_0, x_1, \ldots, x_c\}$ in $U_{c,d}$ such that $g(x_1), \ldots, g(x_c)$ have been already defined but not $g(x_0)$. We add new nodes and edges as instructed and define $f(e_0) = x_0$ and $g(x_0) = n_0$. Let $\phi_1, \ldots, \phi_c$ be the global coding vectors assigned on the in-edges of $g(x_1), \ldots, g(x_c)$. As $\phi_1, \ldots, \phi_c$ forms a basis in $F^c$, we can assign any of the $\lvert F \rvert ^c$ possible global coding vectors to edge $e_0$. Note $c\leq k < d$. Assume that global coding vectors $\phi_1, \ldots, \phi_k$ have been chosen and let $S_k$ be the set of nonzero vectors $\phi$ for which there exists $\phi_{i_1}, \ldots, \phi_{i_{c-1}}$ such that $\phi \in \Span(\phi_{i_1}, \ldots, \phi_{i_{c-1}})$. Since there are $\binom{k}{c-1}$ ways to choose $\phi_{i_1}, \ldots, \phi_{i_{c-1}}$, which forms an independent set, and $\lvert F \rvert ^{c-1} - 1$ nonzero vectors in $\Span(\phi_{i_1}, \ldots, \phi_{i_{c-1}})$, it follows that $\lvert S_k \rvert \leq \binom{k}{c-1}\left(\lvert F \rvert^{c-1} - 1 \right)$. There are $\lvert F \rvert^c -1$ nonzero vectors in $F^c$ and exactly $\lvert S_k \rvert$ of them does not work as $\phi$ to maintain (\ref{prop:u1}). Since $\lvert F \rvert \geq \binom{d-1}{c-1}$ and $x < \frac{x^c-1}{x^{c-1}-1}$ when $x>1$ and $c\geq 2$,
\begin{equation*}
\binom{k}{c-1} \leq \binom{d-1}{c-1} \leq \lvert F \rvert < \frac{\lvert F \rvert ^c -1}{\lvert F \rvert ^{c-1} -1}.
\end{equation*}
Then
\begin{equation*}
\lvert S_k \rvert \leq \binom{k}{c-1}\left(\lvert F \rvert^{c-1} - 1 \right) < \lvert F \rvert ^c -1.
\end{equation*}

Hence we can choose a global coding vector $\phi$ for edge $e_0$ such that there exists no $\phi_{i_1}, \ldots, \phi_{i_{c-1}}$ such that $\phi \in \Span(\phi_{i_1}, \ldots, \phi_{i_{c-1}})$; that is $\phi$ together with all the global coding vectors already defined satisfies (\ref{prop:u1}). After all the iterations of Step 2, we have $\phi_1, \ldots, \phi_d$ such that any $c$ of them forms a basis in $F^c$. 

We show that demands of any receiver node to be created in Steps 3-4 can be satisfied. In Step 3, we choose a circuit $\{x_0, \ldots, x_c\}$ where $g(x_0)$ is a source node with message $m_0$. We choose appropriate edge functions on edges $e_1, \ldots, e_c$ so that the new node $y$ receives coded messages corresponding to the global coding vectors of in-edges of $g(x_1), \ldots, g(x_c)$. Since the global coding vectors form a basis of $F^c$, we can recover all the standard basis vectors $\tilde{e}_1, \ldots, \tilde{e}_c$ at node $y$ and therefore all the network messages $m_1,\ldots, m_c$ and in particular $m_0$. Hence the demands of any receiver node to be created in Step 3 can be satisfied. By similar argument, the same follows for receiver nodes to be created in Step 4. 
\end{IEEEproof}

\begin{theorem}\label{thm:graphic}
Any matroidal network constructed from a graphic matroid is scalar-linearly solvable over sufficiently large finite field of characteristic 2. 
\end{theorem}
\begin{IEEEproof}
The proof idea is the same as before. Let $G=(V,E)$ be an undirected graph with node set $V$ and edge set $E$ and let $\MMM(G)$ be the corresponding graphic matroid. It suffices to consider the case where $G$ is connected and has no self-loops. If $G$ has several connected components, the resulting network $\NNN$ is a combination of smaller networks constructed from each component. A self-loop is a single-edge cycle and does not change $\NNN$ in any significant way in Steps 1-4, so we can just ignore self-loops. It follows that there are exactly $\lvert V \rvert -1$ messages in $\NNN$.

Instead of constructing the global coding vectors with respect to the network $\NNN$, we construct them directly on the edges of $G$. This is because defining $g(x)$ for some $x\in E$ and assigning a global coding vector for the in-edge of $g(x)$  (or the auxiliary edge for the source $g(x)$) corresponds to associating the same global coding vector to the edge $x$. To make sure that all demands of any receiver node to be created in Steps 3-4 can be satisfied, we construct global coding vectors on the edges such that the following conditions are satisfied:
\begin{enumerate}
	\item If $C=\{e_0, e_1, \ldots, e_j\}$ is a cycle in $G$ with $g(e_0)$ being a source node with message $m_0$, then the standard basis vector $\tilde{e}$ corresponding to $m_0$ can be recovered scalar-linearly from the global coding vectors assigned to $e_1, \ldots, e_j$.
	\item If $B=\{e_1, \ldots, e_{\lvert V\rvert -1}\}$ is a spanning tree in $G$, then all the standard basis vectors $\tilde{e}_1, \ldots, \tilde{e}_{\lvert V\rvert -1}$ corresponding to the messages can be recovered scalar-linearly from the global coding vectors assigned to $e_1, \ldots, e_{\lvert V \rvert -1}$.
\end{enumerate}

We choose a finite field $F$ of characteristic 2 that is sufficiently large enough so that each symbol in the alphabet $\AAA$ can be represented. Note that $F=\FF_{2^l}$ for some sufficiently large $l$. In Step 1, we choose a spanning tree $T =\{h_1, \ldots, h_{\lvert V \rvert -1}\}$ and assign the standard basis vectors $\tilde{e}_1, \ldots, \tilde{e}_{\lvert V\rvert -1}$ as global coding vectors on edges $h_1, \ldots, h_{\lvert V \rvert -1}$. In Step 2, we repeatedly find a cycle $C = \{e_0, e_1, \ldots, e_j\}$ and assign a global coding vector to edge $e_0$ using global coding vectors of $e_1, \ldots, e_j$. We prove by induction on the number of edges that have been assigned a global coding vector that we can assign $\phi$ to each new edge $e_0$ in Step 2 such that 
\begin{equation}\label{prop:g1}
\phi_{e_0} = \tilde{e}_{i_1} + \cdots + \tilde{e}_{i_k},
\end{equation}
where the corresponding edges $h_{i_1}, \ldots, h_{i_k}$ of $T$ and edge $e_0$ forms an unique cycle when $e_0$ is added to $T$. Clearly, the first edge $e_0$ to be processed in Step 2 satisfies (\ref{prop:g1}) since $e_1, \ldots, e_j$ are precisely the edges of the spanning tree $T$. Assume that the edges have been assigned global coding vectors in this way and consider a new cycle $C= \{e_0, e_1, \ldots, e_j\}$ with only $e_0$ unassigned. Let cycle $C' = \{e_0, h_{i_1}, \ldots, h_{i_k}\}$ be the unique cycle induced by adding edge $e_0$ to $T$, ordered so that one edge leads to the next. Let $v_0, \ldots, v_j$ be the nodes on the cycle $C$ and assume $e_i$ connects $v_i$ to $v_{i+1}$, where $j+1 \equiv 0$. Let $\phi_{e_i}$ be the global coding vector assigned to edge $e_i$.

For each edge $e_i$ in $C$, there is exactly one path from $v_i$ to $v_{i+1}$ in the spanning tree $T$, say $P_i$. The path $P_0$ for $e_0$ is exactly $(h_{i_k}, \ldots, h_{i_1})$. Then the concatenation of paths $P_0 + P_1 + \cdots + P_j$ is a cycle using edges of $T$. If an edge appears twice in a row in $P_0 + P_1 + \cdots + P_j$, we remove the occurrences and still obtain a cycle in $T$. Since there exists no cycle in the spanning tree $T$, the cycle $P_0 + P_1 + \cdots + P_j$ must reduce to the empty cycle after such removals. It follows that $P_1 + \cdots + P_j$ reduces to $(h_{i_1}, \ldots, h_{i_k})$ to cancel out path $P_0$. This relation combined with the fact that $1+1 = 0$ in field $F$ and that (\ref{prop:g1}) holds for the global coding vectors that have been assigned so far implies that
\begin{equation*}
\phi_{e_1} + \cdots + \phi_{e_j} = \tilde{e}_{i_1} + \cdots + \tilde{e}_{i_k}. 
\end{equation*}
We let $\phi = \phi_{e_1} + \cdots + \phi_{e_j}$ to be the global coding vector for edge $e_0$.

We show that this construction of global coding vectors on edges of $G$ satisfies the conditions mentioned in the beginning. In Step 3, if $C=\{e_0, e_1, \ldots, e_j\}$ is a cycle with $e_0$ being one of the edges of $T$, say $h_i$, (which is equivalent to saying $g(e_0)$ is a source node) then the standard basis vector $\tilde{e}_i$ can be recovered by $\tilde{e}_i = \phi_{e_0} = \phi_{e_1} + \cdots + \phi_{e_j}$. In Step 4, assume $B=\{e_1, \ldots, e_{\lvert V \rvert -1 }\}$ is a spanning tree. For each $i=1, \ldots, \lvert V\rvert -1$, if $(h_i, e_{i_1}, \ldots, e_{i_k})$ is the unique cycle induced by adding $h_i$ to $B$ then $\tilde{e}_i = \phi_{e_{i_1}} + \cdots + \phi_{e_{i_k}}$. Hence we can deduce all the network messages. 
\end{IEEEproof}

Note that from Theorems \ref{thm:construction} and \ref{thm:converse} we get a method for constructing scalar-linearly solvable networks: pick any representable matroid over a finite field $F$ and construct a matriodal network $\NNN$ using Theorem~\ref{thm:construction}.

\section{Conclusion}
In this paper, we showed that any matriodal network associated with a representable matroid over a finite field is scalar-linearly solvable. Combined with an earlier result of Dougherty et al., it follows that a network is scalar-linearly solvable if and only if it is a matroidal network associated with a representable matroid over a finite field. Using the theorem implicitly, we showed that matroidal networks constructed from uniform and graphic matroids are scalar-linearly solvable. As result, we got a set of scalar-linearly solvable networks that might be different from those networks we already know. We also noted a method for generating scalar-linearly solvable networks from representable matroids over a finite field that might have a good chance of being not trivially solvable by routing.

%


\section*{Acknowledgment}
This material is based upon work supported by the Air Force Office of  Scientific
Research (AFOSR) under award No. 016974-002.



\bibliographystyle{IEEEtran}
\bibliography{IEEEabrv,mybib}
%

%
%

\end{document}